\documentclass[submission,Phys]{SciPost}
\usepackage[T1]{fontenc}
\usepackage{lmodern}
\usepackage{amsmath,amssymb,amsthm,mathtools,mathrsfs}
\usepackage{microtype}
\usepackage{dsfont}
\usepackage{booktabs}
\hypersetup{
  colorlinks=true,
  linkcolor=scipostdeepblue,
  citecolor=scipostdeepblue,
  urlcolor=scipostdeepblue,
  pdftitle={Effective form factors at a quadratic saddle: the Weber sigmoid and the Barnes prefactor}
}
\newcommand{\eprint}[2][]{\href{https://arxiv.org/abs/#2}{\nolinkurl{arXiv:#2}}}
\allowdisplaybreaks[2]
\numberwithin{equation}{section}

\newcommand{\R}{\mathbb{R}}

\begin{document}
\begin{center}
{\Large\bfseries Finite temperature Fredholm determinants for dynamical correlation functions}
\end{center}

\begin{center}
Oleksandr Gamayun
\end{center}

\begin{center}
London Institute for Mathematical Sciences, Royal Institution, 21 Albemarle, London, W1S 4BS, UK
\\
 og@lims.ac.uk
\end{center}


\begin{center}\today \end{center}

\section*{Abstract}
{\bfseries

We study the long-time, large-distance asymptotics of finite-temperature
dynamical Fredholm determinants using effective form factors. The
method expresses these asymptotics as a functional of an effective
phase shift. We determine its two branches on either side of a
quadratic stationary point and the transition profile joining them.
The profile is expressed through parabolic-cylinder functions. Its
connection to the high-temperature Gaussian $\beta$ ensemble fixes
the finite normalization, in agreement with an independent
Riemann--Hilbert analysis. We also show that the leading exponential
asymptotics suffice to recover the Euler-scale prediction of generalized
hydrodynamics for the dynamical structure factor of hard rods.
\par}

\vspace{10pt}
\noindent\rule{\textwidth}{1pt}
\tableofcontents
\thispagestyle{fancy}
\noindent\rule{\textwidth}{1pt}
\vspace{10pt}


\section{Introduction}

Real-time correlation functions connect the microscopic excitations of
quantum many-body systems to experimental observables via various response functions 
\cite{AltlandSimons2023,Sachdev2011}. For example, density correlations
determine the dynamical structure factor measured by Bragg spectroscopy
in ultracold gases \cite{Fabbri2015}, while spin correlations can be
probed through nuclear magnetic resonance relaxation rates
\cite{Steinberg2019}. The momentum and frequency dependence of these
correlations reveals the excitation spectrum, while their long-distance
and long-time behaviour characterizes collective processes such as
transport and thermal relaxation. A quantitative description of these
processes requires knowledge of the correlation amplitudes, in addition
to the decay rates and scaling exponents.

At finite temperature, or more generally at finite entropy states, the direct computation of the real time correlation functions becomes challenging, as it involves addressing the dynamics of a
macroscopically populated background. Even for one-dimensional quantum systems, numerical
real-time evolution is limited by entanglement growth 
\cite{CalabreseCardy2005,Paeckel2019}. Some features of the correlation functions of the local operators on large
space and time scales, can be addressed by the (generalized) 
hydrodynamics, in which case the complexity is typically
reduced to much fewer degrees of freedom \cite{DeNardis2022}.  
In integrable systems, generalized hydrodynamics describes the evolution of
local quasiparticle distributions
\cite{CastroAlvaredoDoyonYoshimura2016,BertiniColluraDeNardisFagotti2016}
and determines Euler-scale correlations of conserved densities
\cite{Doyon2018}. These results provide a natural starting point for
long-time asymptotics. To go beyond this scale, or to treat observables
whose leading behaviour is not fixed by hydrodynamic projection, one
must recover information about microscopic matrix elements and perform the corresponding spectral sums.

An important limit of correlation functions in integrable models is the strong-coupling limit. In this limit, correlation functions can often be expressed as Fredholm determinants. Their asymptotics can then be studied through the associated Riemann--Hilbert problem. Nonlinear steepest descent provides a systematic method for this analysis \cite{DeiftZhou1993,DeiftItsZhou1997}, with the generalized sine kernel furnishing an important example \cite{KitanineEtAl2009}. Besides the leading exponential behaviour, this method can resolve power laws and finite normalization factors. Nevertheless, constructing the required transformations and local solutions for a dynamical kernel can involve a substantial calculation. It is therefore useful to seek a complementary description that makes the same asymptotic structure transparent from the form factor expressions.
 
Numerous attempts have been made to develop heuristic methods for extracting at least the leading asymptotic behavior from form factor series. These methods have proved effective in particular regimes. Semiclassical descriptions of low-temperature Ising-chain dynamics \cite{SachdevYoung1997} and of relaxation near two-dimensional quantum-critical points \cite{Sachdev1999} connected long-time decay to the motion and scattering of thermal excitations. Subsequent form factor resummations yielded large-distance and long-time asymptotics in integrable models \cite{AltshulerKonikTsvelik2006}, low-temperature NMR response \cite{Steinberg2019}, and thermal and post-quench correlations in the transverse-field Ising chain \cite{GranetFagottiEssler2020}. More recently, determinant representations have enabled detailed studies of dynamical correlations in Ising field theory \cite{CsepanyiKormos2024}. Although these methods apply in different settings, they show how an appropriate organization of the spectral expansion can make collective dynamics accessible without evaluating its terms separately.

A systematic microscopic resummation is also available at zero
temperature for a class of critical models. There, form-factor methods
recover the large-distance asymptotics and amplitudes of static
correlations \cite{KitanineEtAl2011,KozlowskiMaillet2015}, and extend to
dynamical correlations with contributions from both low-energy
excitations and stationary points of the dispersion
\cite{KitanineEtAl2012}. This connects the microscopic description to
the nonlinear Luttinger-liquid picture
\cite{ImambekovSchmidtGlazman2012}.

Inspired by these developments we introduced the effective form factor (EFF) approach
\cite{GamayunIorgovZhuravlev2021,ZhuravlevEtAl2022,ChernowitzGamayun2022}
with the aim of extending the same degree of control 
to the correlation functions in states with finite entropy. 
Its basic idea is to replace the original form factors by a solvable
free-fermionic family whose phase shift and weight are chosen to match
the relevant thermodynamic data. The auxiliary form factor series can
then be summed explicitly, leaving a functional of the effective phase
shift. This separates the resummation from the matching problem: the
former is exact within the auxiliary family, while the latter determines
whether it reproduces the desired asymptotic regime. For static
determinants, this construction also has a direct interpretation through
an exactly solvable auxiliary Riemann--Hilbert problem
\cite{gamayun_finite-temperature_2025}. The effective determinant can thus be
introduced and evaluated directly in the thermodynamic limit.

In the dynamical case, the remaining matching problem contains an
essential local feature. At fixed ray $x/t$, the derivative of the
oscillatory phase changes sign at a stationary momentum. The Cauchy
integrals entering the kernel consequently select different effective
phase shifts on its two sides. These outer solutions generally have a
finite mismatch, so they must be joined within a neighbourhood of width
$t^{-1/2}$ for a quadratic saddle. Earlier EFF calculations identified
this smoothing mechanism and obtained the exponential rate and the
additional power of time \cite{ChernowitzGamayun2022}.  We show in this work that these outer
asymptotics already suffice to recover the generalised hydrodynamic prediction for the
Euler-scale dynamical structure factor of hard rods
\cite{GamayunPanfil2026}. The finite amplitude of the determinant,
however, depends on the transition profile and remains undetermined
if that profile is chosen arbitrarily. The recent Riemann--Hilbert
analysis of G\"ohmann, Kozlowski, and Minin \cite{Minin2026} gives
renewed motivation to resolve this normalization problem directly
within the EFF construction.

In this work, we determine the transition profile by formulating and solving the sewing conditions in the $t^{-1/2}$ zone around the quadratic saddle. In doing so, we encounter a Riccati equation for the local resolvent similar to that arising in beta ensembles. The profile is expressed in terms of parabolic-cylinder functions and reduces to a complex error function in the small-jump limit. Its dependence on the microscopic state enters through the phase-shift jump $\Delta$ at the saddle.

After a contour rotation, the derivative of the profile becomes the Gauss--Wigner crossover density of the high-temperature Gaussian beta ensemble \cite{AllezBouchaudGuionnet2012}. In this ensemble, the entropy of the density contributes at the same order as the confining potential and logarithmic interaction. The resulting variational principle reproduces the local resolvent equation, while the exact Gaussian partition function determines the value of the functional at its minimizing density. This connection fixes the previously undetermined local constant and yields the Barnes factor $(2\pi)^{\Delta/2}G(1-\Delta)$ in our conventions, in agreement with the amplitude obtained by Minin et al. through a full Riemann--Hilbert analysis \cite{Minin2026}. It thus explains how the finite normalization arises from the saddle regularization.

The manuscript is organized as follows. Section~\ref{sec:eff}
introduces the effective determinant, derives the outer phase
shifts for a momentum-dependent phase, and shows how they yield
the Euler-scale asymptotics of the hard-rod dynamical structure
factor. Section~\ref{sec:sewing} formulates and solves the local
sewing problem. Section~\ref{sec:rmt} uses the Gaussian ensemble
to evaluate the finite constant. Section~\ref{sec:rhp} relates
exact kernel matching to the standard Riemann--Hilbert problem.
The appendices prove the exact effective-determinant identity
and give an alternative Fourier construction of the Weber profile.

\section{Effective form factors for a momentum-dependent phase}
\label{sec:eff}

We consider the long-time, large-distance asymptotics of the following time-dependent generalized sine-kernel Fredholm determinant
\begin{equation}\label{t0}
    \mathcal{D}_\nu(x,t)=\det(1+\hat K),\qquad
    K(q,p)=
    \frac{e_+(p)e_-(q)-e_-(p)e_+(q)}{p-q},
\end{equation}
with $x/t$ held fixed. The operator $\hat K$ acts on $L^2(\mathbb R)$, and
\begin{equation}\label{eq:epem}
    \begin{split}
        e_-(q) &= \sqrt{\rho(q)}\sin[\pi \nu(q)]e^{\frac{i x \mathfrak{p}(q)}{2}},\qquad x\mathfrak{p}(q) = xq - q^2 t/2\\
        e_+(q) &= \frac{e_-(q)}{4\pi^2 \sin[\pi \nu(q)]^2}\int dk \left[\frac{1-e^{2\pi i \nu(k) }}{k-q+i0}+\frac{1-e^{-2\pi i \nu(k)}}{k-q-i0}\right]e^{-ix\mathfrak{p}(k)}.
    \end{split}
\end{equation}
This determinant provides a setting in which to examine the asymptotics of dynamical correlation functions. The function $\rho$ specifies the occupation profile of the state; for a thermal state, it has the Fermi--Dirac form
$\rho(k)\sim[e^{(k^2/2-\mu)/T}+1]^{-1}$.
The phase $\nu(k)$ encodes scattering information. In particular,
$\nu=1/2$ is relevant to the impenetrable-boson case
\cite{Lenard1964Momentum}. For constant $\nu$, the determinant also
appears as a building block of the dynamical structure factor of
quantum hard rods \cite{GamayunPanfil2026}. There, the physical
correlator is obtained by integrating the fixed-$\nu$ result with the
appropriate weight. This integration can be also understood as averaging over statistics \cite{PhysRevA.109.012209}.

Compared with the kernel introduced in Ref.~\cite{Minin2026}, we have made minor changes of notation to remain consistent with Refs.~\cite{GamayunPanfil2026,ChernowitzGamayun2022}. We also set the auxiliary function $g(q)=0$ and choose their dispersion $u(\lambda|\lambda_0)$ to be quadratic, simplifying the analysis while retaining the dynamical stationary point of interest.

Our strategy for finding the asymptotic expansion of $\mathcal{D}_\nu(x,t)$ is to split the kernel as $\hat{K} =  \hat{\mathcal{V}}+\hat{\mathcal{W}}$. The determinant then factorizes as
\begin{equation}\label{eq:kernel-factorization}
    \mathcal{D}_\nu(x,t) =  \tau(x,t) \det\left(1+ (1-\hat{\mathcal{R}})\hat{\mathcal{W}}\right)
\end{equation}
with
\begin{equation} 
\tau(x,t) = \det(1 + \hat{\mathcal{V}}) , \qquad (1-\hat{\mathcal{R}})(1+\hat{\mathcal{V}})=1.
\end{equation}
The choice of $\hat{\mathcal{V}}$ is guided by two requirements: $\tau(x,t)$ should be explicitly computable, and $\hat{\mathcal{V}}$ should approximate the original kernel closely enough to reproduce its leading asymptotics as $x,t\to\infty$.

Microscopic form factors \cite{ChernowitzGamayun2022} suggest a kernel of the form
\begin{equation}\label{tt}
\mathcal{V}(q,p) = \frac{v_+(p)v_-(q)-v_-(p)v_+(q)}{p-q}
\end{equation}
with
\begin{equation}\label{eq:epem1}
    \begin{split}
        v_-(q) &= \sqrt{r(q)} \exp\left(\frac{i x \mathfrak{p}(q)}{2}\right),\\
        v_+(q) &= \frac{v_-(q)}{4\pi^2}\int \frac{dk}{r(k)} \left[\frac{1-e^{2\pi i \eta(k) }}{k-q+i0}+\frac{1-e^{-2\pi i \eta(k)}}{k-q-i0}\right]e^{-ix\mathfrak{p}(k)}.
    \end{split}
\end{equation}
For sufficiently regular functions $\eta(k)$ and $r(k)$\footnote{In particular, we require that combinations $
1+\rho(k)(e^{\pm2\pi i\nu(k)}-1)
$ do not vanish”. We also require zero index for $\eta'(k)$ over the real line i.e. \(
\eta(+\infty)-\eta(-\infty)
=\int_{\mathbb R}\eta'(k)dk=0
\).}, its determinant has the explicit expression
\begin{equation}
\ln\tau(x,t)
=\int_\R\eta(k)
\left[i(x-tk)+\frac{r'(k)}{r(k)}\right]d k
+\iint_{\R^2}\eta'(p)\eta'(q)\ln|p-q|\,d p\,d q.
\label{eq:eff-functional}
\end{equation}
This expression can also be derived directly from the Fredholm determinant, without summing microscopic form factors. We gave such a derivation for the static case and a lattice system in \cite{gamayun_finite-temperature_2025}. For completeness, we outline the corresponding derivation here in Appendix~\ref{appA}.

We now seek an effective kernel $\mathcal{V}$ in \eqref{tt} that asymptotically matches $\mathcal{K}$ in \eqref{t0}. For $q$ away from the critical point $k_0= x/t$, the oscillatory Cauchy factor obeys
\begin{equation}
    \lim\limits_{x\to+\infty} \frac{1}{2\pi i}\frac{e^{-i\mathfrak{p}(k) x}}{k-q+i0} = \begin{cases} 
      -e^{-i\mathfrak{p}(q) x}\delta(k-q) , &  \mathfrak{p}'(q) > 0 \\
      0,  &  \mathfrak{p}'(q) < 0,
   \end{cases}
\end{equation}
This identifies
\begin{equation}
    \sqrt{r(q)} = \sqrt{\rho(q)} \sin[\pi \nu(q)] 
\end{equation}
and gives two matching conditions for the effective phase
\begin{equation}\label{f}
\begin{cases}
e^{2\pi i \eta(k)}-1
=\rho(k)(e^{2\pi i \nu(k)}-1),&k<k_0,\\[6pt]
e^{-2\pi i \eta(k)}-1
=\rho(k)(e^{-2\pi i \nu(k)}-1),&k>k_0.
\end{cases}
\end{equation}
These conditions define separate left and right branches. Combining them gives the discontinuous effective phase
\begin{equation}\label{sharp}
\eta(k) = \eta_\ell (k) + (\eta_r(k) - \eta_\ell(k)) \theta(k-k_0) \equiv \eta_0(k)
\end{equation}
where $\theta$ is the step function and, by \eqref{f},
\begin{equation}
    \eta_\ell(k) = \frac{\ln\left[1+\rho(k) (e^{2\pi i \nu(k)}-1)\right]}{2\pi i} ,\qquad  \eta_r(k) = -\frac{\ln\left[1+\rho(k) (e^{-2\pi i \nu(k)}-1)\right]}{2\pi i }
\end{equation}
For real $\rho$ and $\nu$, the jump at $k_0$ is
\begin{equation}
    \Delta \equiv  \eta(k_0+0)-\eta(k_0-0) = \frac{i}{2\pi}\ln \left|1+\rho(k_0) (e^{-2\pi i \nu(k_0)}-1)\right|^2 \neq 0
\end{equation}
The jump makes the double integral in \eqref{eq:eff-functional} divergent. It must therefore be smoothed on the scale $1/\sqrt{t}$, as suggested in Ref.~\cite{ChernowitzGamayun2022}. In place of the sharp phase \eqref{sharp}, we take
\begin{equation}
    \eta(k) = \eta_\ell(k) + (\eta_r(k) - \eta_\ell(k)) S_\Delta (\sqrt{t}(k-k_0))
\end{equation}
where $S_\Delta$ is a smooth interpolating function satisfying $S_\Delta(-\infty)=0$ and $S_{\Delta}(+\infty)=1$.

Substituting this phase into the exact expression \eqref{eq:eff-functional} gives the asymptotic form\footnote{For the detailed derivation see Eqs. (137) - (145) in \cite{ChernowitzGamayun2022}}
\begin{multline}\label{as11}
\ln\tau(x,t) 
 \approx  C_0 -\frac 1 2\Delta^2 \ln (it)  +\int dk\,\eta_0(k)\left(i(x-tk)+\frac{r'(k)}{r(k)}\right) \\ 
 -\frac{1}{2}  \int_\R dp \int_\R dq\frac{\eta_0(q)\dot{\eta}_0(p)-\dot{\eta}_0(q)\eta_0(p)}{q-p} +\Delta \int_\R dk\, \dot{\eta}_0(k) \ln|k-k_0|.    
\end{multline}
Here $\approx$ means that we have dropped terms $O(1/\sqrt{t})$; $\eta_0$ is defined in \eqref{sharp}, while $\dot{\eta}_0$ denotes the derivative taken separately on its two smooth branches
\begin{equation}
\dot{\eta}_0(k) = \eta'_\ell (k) + (\eta'_r(k) - \eta'_\ell(k)) \theta(k-k_0).
\end{equation}
The dependence on the interpolating function $S_\Delta$ is contained in the constant $C_0$:
\begin{equation}
      C_0=\frac{i\pi \Delta^2}{4}-i\Delta \int_\R z\left( S_\Delta(z) - \theta(z) \right)dz+\Delta^2\iint_{\mathbb{R}^2}\,S_\Delta'(p) S_\Delta'(q) \ln\left|p-q\right| dp  \,dq 
\end{equation}
Equivalently, integration by parts gives
\begin{equation}\label{c0}
      C_0=\frac{i\pi \Delta^2}{4}+\frac{i\Delta}{2} \int_\R z^2 S'_\Delta(z)dz+\Delta^2\iint_{\mathbb{R}^2}\,S_\Delta'(p) S_\Delta'(q) \ln\left|p-q\right| dp  \,dq 
\end{equation}
The first term reflects our convention of writing $\ln(it)$ in \eqref{as11}. The second was absent in Ref.~\cite{ChernowitzGamayun2022} because of the assumptions made there about the regularization. Since the interpolating function was not determined, $C_0$ also remained undetermined.

The full Riemann--Hilbert analysis of Ref.~\cite{Minin2026} identifies the missing prefactor as
\begin{equation}\label{prefactor}
    e^{C_0} = (2\pi)^{\Delta/2}G(1-\Delta) 
\end{equation}
where $G$ is the Barnes $G$-function. This prefactor coincides with the constant $\tau_\alpha$ in the large-gap asymptotics of the hard-edge Bessel-kernel determinant when $\alpha=-\Delta$ \cite{TracyWidom1994,Ehrhardt2010}. For $\Delta=-N$, it also has the form of a suitably normalized $N$-eigenvalue Gaussian unitary ensemble partition function. In the next sections, we derive and solve a condition that fixes $S_\Delta$, show that \eqref{c0} yields \eqref{prefactor}, and establish a connection with the high-temperature limit of the Gaussian ensemble.

{\bf Remark.} Overall the asymptotic expression for the determinant can be written as
\begin{equation}
    \mathcal{D}_{\nu}(s,t) \approx \mathcal{A}(\nu, s/t,t) \,
    e^{i\int dk\,\eta_0(k)(s-tk)} .
\end{equation}
Here $\eta_0$ is defined in \eqref{sharp}, while $\mathcal{A}$ contains $C_0$ and the remaining terms of \eqref{as11}. The leading connected correlation on the Euler scale is governed by small $\nu$. At this order, the variation of $\mathcal{A}$ can be neglected and we may set $\mathcal{A}\approx\mathcal{A}(0,s/t,t)=1$.

Let us illustrate this for the hard rods gas\cite{GamayunPanfil2026}. The density correlation function has the representation
\begin{equation}\label{Sxt}
    S(x,t) =\int\limits_{-\infty}^{\infty} \frac{dP}{2\pi} \,
    \frac{\sqrt{K}P^2e^{-iPx}}{\left(2\sin(\pi\nu)\right)^2}
    \int\limits_{-\infty}^{\infty} ds\,e^{isP}\mathcal{D}_\nu(s,t).
\end{equation}
Here $K=(1-\rho_0a)^2$ and $\nu=aP/(2\pi)$, where $\rho_0=N/L$ is the mean density and $a$ is the rod length. We assume a symmetric filling $\rho(k)$, normalized by
$\int\rho(k)dk=2\pi\rho_0/\sqrt K$. Define
\begin{equation}
    B_t(s)=\int \rho(k)(1-\rho(k))|s-kt|\frac{dk}{4\pi}.
\end{equation}
For constant $\nu$, the small-$\nu$ expansion of the phase gives
\begin{equation}
    i\int dk\,\eta_0(k)(s-tk)
    =\frac{2\pi i\rho_0}{\sqrt K}\nu s
     -(2\pi\nu)^2B_t(s)+O(t\nu^3)
\end{equation}
on a fixed ray $s/t$. Since $\nu=aP/(2\pi)$, the linear term combines with $e^{isP}$ in \eqref{Sxt}. Expanding at small $P$, we obtain
\begin{multline}\label{ala}
    S(x,t)\approx
    \int\limits_{-\infty}^{\infty}\frac{\sqrt K\,dP}{2\pi a^2}
    \int\limits_{-\infty}^{\infty}ds\,
    e^{i(s/\sqrt K-x)P}e^{-a^2P^2B_t(s)}
    \\
    =
    \int\limits_{-\infty}^{\infty}\frac{K\,ds}{2\pi a^2}
    \int\limits_{-\infty}^{\infty}dP\,e^{isP}
    \left[1-a^2P^2B_t\bigl(\sqrt K(s+x)\bigr)+\cdots\right].
\end{multline}
In the second line we shifted the integration variable according to
$s\to\sqrt K(s+x)$.

The first term in the square brackets is independent of time. Its coefficient is not fixed reliably by this small-$P$ expansion, which is non-uniform under the integral over the unbounded variable $s$. The disconnected background is $\rho_0^2$; equivalently, it can be removed by considering $S(x,t)-S(x,0)$. The term proportional to $P^2$ determines the leading connected correlation on a nonzero Euler ray. Using the Fourier representation of the second derivative of the delta function, we find
\begin{equation}
    S_c(x,t)\equiv S(x,t)-\rho_0^2
    \sim K\partial_s^2B_t(\sqrt K\,s)\Big|_{s=x}
    =\frac{K^2}{2\pi t}\,
       \rho\!\left(\frac{\sqrt K\,x}{t}\right)
       \left[1-\rho\!\left(\frac{\sqrt K\,x}{t}\right)\right].
\end{equation}

This agrees with the Euler-scale prediction of generalized hydrodynamics for a homogeneous state \cite{Doyon2018}. A mode with rapidity $k$ propagates at its effective velocity; its contribution to the density correlation is weighted by the occupation variance $\rho(k)[1-\rho(k)]$ and the square of its dressed particle charge. For the symmetric hard-rod state and the dispersion $k^2/2$, these quantities are
\begin{equation}
    \rho_{\rm p}(k)=\frac{\sqrt K}{2\pi}\rho(k),
    \qquad
    1^{\rm dr}(k)=\sqrt K,
    \qquad
    v^{\rm eff}(k)=\frac{k}{\sqrt K}.
\end{equation}
The GHD projection formula therefore reads
\begin{equation}
    S_c(x,t)
    \sim\int dk\,\rho_{\rm p}(k)[1-\rho(k)]
       [1^{\rm dr}(k)]^2
       \delta\!\left(x-v^{\rm eff}(k)t\right),
\end{equation}
and reproduces the result above. Ref.~\cite{GamayunPanfil2026} uses the dispersion $k^2$, for which $v^{\rm eff}(k)=2k/\sqrt K$.

Hydrodynamic correlations can also be obtained from thermodynamic
form factors. For the Lieb--Liniger model this approach was introduced in 
\cite{DeNardisPanfil2016}. An axiomatic formulation of thermodynamic
form factors for integrable quantum field theories was developed
in Ref.~\cite{CortesCuberoPanfil2019}. Within this framework, the
contribution of a single particle--hole pair reproduces the
Euler-scale correlations \cite{CortesCuberoPanfil2020}, while
contributions from two particle--hole pairs give access to
correlations at diffusive order \cite{Panfil2021}. We expect that a more detailed asymptotic analysis of the exact
representation \eqref{Sxt} will also give access to corrections
beyond the Euler scale.
 
\section{Saddle point regularization}
\label{sec:sewing}

Let us now focus on momenta close to $k_0$. We introduce the scaling
$k=k_0+s/\sqrt{t}$ in the integrals for $e_+$ and $v_+$ in
Eqs.~\eqref{eq:epem} and \eqref{eq:epem1}, respectively, and evaluate
both functions at $q=k_0+z/\sqrt{t}$. We also replace the smooth
functions in the integrands by their values at $k_0$. This gives
\begin{equation}
    e_+(q) \sim \frac{e_-(q)e^{-ix\mathfrak p(k_0)}}{4\pi^2 \sin[\pi \nu(k_0)]^2}\int_{\R} ds  \left[\frac{1-e^{2\pi i \nu(k_0)}}{s-z+i0}+\frac{1-e^{-2\pi i \nu(k_0)}}{s-z-i0}\right]e^{is^2/2}
\end{equation}
and
\begin{equation}
    v_+(q) \sim \frac{e_-(q)e^{-ix\mathfrak p(k_0)}}{4\pi^2  \sin[\pi \nu(k_0)]^2}\int_{\R} \frac{ds}{\rho(k_0)}  \left[\frac{1-e^{2\pi i \eta_\ell(k_0)}w(s)}{s-z+i0}+\frac{1-e^{-2\pi i \eta_\ell(k_0)}w(s)^{-1}}{s-z-i0}\right]e^{is^2/2},
\end{equation}
where $w(s)=\exp(2\pi i\Delta S_\Delta(s))$. For $z=O(1)$,
the condition $e_+=v_+$ in fact gives two relations. Indeed,
applying the upper and lower Cauchy projections separates the matching relation into two conditions
\begin{equation}\label{cond1}
    \int_{\R} \frac{(w(s)-1)e^{is^2/2}}{s-z+i0}ds =0,\qquad
    \int_{\R} \frac{(e^{2\pi i \Delta }w(s)^{-1}-1)e^{is^2/2}}{s-z-i0}ds =0.
\end{equation}
We can first try to solve these conditions perturbatively in $\Delta$. At
first order, we have
\begin{equation}
    w(s) = 1 + \Delta f_1(s)+o(\Delta),\qquad
    w(s)^{-1} = 1 - \Delta f_1(s)+o(\Delta).
\end{equation}
The first condition gives
\begin{equation}
    \int_{\R} \frac{f_1(s)e^{is^2/2}}{s-z+i0}ds=0,
    \qquad\Rightarrow\qquad
    \int_{\R} \frac{f_1(s)e^{is^2/2}}{s-z-i0}ds
       =2\pi i e^{iz^2/2}f_1(z).
\end{equation}
At the same order, the second condition reads
\begin{equation}
    \int_{\R}
    \frac{(2\pi i-f_1(s))e^{is^2/2}}{s-z-i0}ds=0.
\end{equation}
Together, they yield
\begin{equation}
    f_1(s)=\pi i\left(1+
    {\rm Erf}\left[\frac{1+i}{2}s\right]\right).
\end{equation}
We could continue the expansion order by order. Under suitable
assumptions on the analytic continuations, however, we can solve
the sewing problem at finite $\Delta$. 

Set
$A(s)\equiv(w(s)-1)e^{is^2/2}$ and
$B(s)\equiv(e^{2\pi i\Delta}w(s)^{-1}-1)e^{is^2/2}$.
The conditions \eqref{cond1} give analytic continuations of $A$
and $B$ into the upper and lower half-planes, respectively. We
introduce the scalar factor
\begin{equation}
\phi(z)=\exp\!\left[
-\Delta \int_{\mathbb R}S_\Delta(s)
\left(\frac{1}{s-z}-\frac{s}{1+s^2}\right)\,ds
\right],\qquad z\in\mathbb C\setminus\mathbb R.
\label{eq:scalar-cauchy-factor}
\end{equation}
The subtraction makes the integral convergent at infinity.
Changing it to another admissible subtraction multiplies $\phi$
by a common, $z$-independent factor. The function $\phi$ is
holomorphic and nonvanishing in each half-plane, and its boundary
values satisfy
\begin{equation}
\phi_-(s)=w(s)\phi_+(s).
\label{eq:scalar-factor-jump}
\end{equation}

Next, define the piecewise holomorphic functions
\begin{align}
a(z)&=
\begin{cases}
e^{iz^2/4}\phi(z),
    &\operatorname{Im}z>0,\\[3pt]
\displaystyle e^{-2\pi i\Delta}
\left(e^{iz^2/4}+e^{-iz^2/4}B(z)\right)\phi(z),
    &\operatorname{Im}z<0,
\end{cases}
\label{eq:scalar-entire-a}\\
b(z)&=
\begin{cases}
\displaystyle
\left(e^{iz^2/4}+e^{-iz^2/4}A(z)\right)\phi(z),
    &\operatorname{Im}z>0,\\[3pt]
e^{iz^2/4}\phi(z),
    &\operatorname{Im}z<0.
\end{cases}
\label{eq:scalar-entire-b}
\end{align}
Substituting the definitions of $A$ and $B$ shows that their
boundary values match: $a_-(s)=a_+(s)$ and $b_-(s)=b_+(s)$.
Since the half-plane definitions have no other finite
singularities, $a$ and $b$ are entire. Moreover,
$b(s)=w(s)a(s)$ for $s\in\R$. Their Wronskian
$W=b'a-a'b$ is therefore entire and can be written as
\begin{equation}
W(z)=
\begin{cases}
\phi(z)^2\bigl(A'(z)-izA(z)\bigr),
&\operatorname{Im}z>0,\\[3pt]
-e^{-2\pi i\Delta}\phi(z)^2\bigl(B'(z)-izB(z)\bigr),
&\operatorname{Im}z<0.
\end{cases}
\label{eq:scalar-wronskian}
\end{equation}

The perturbative solution suggests that $W$ is constant. To
establish this under explicit growth assumptions, suppose that
$A(z),B(z)=O(1)$ and $A'(z),B'(z)=O(|z|)$ as $z\to\infty$
in their respective closed half-planes. The representation of
$\phi$ gives $|\phi(z)|=O(|z|^{\operatorname{Re}\Delta})$,
and consequently
$|W(z)|=O(|z|^{1+2\operatorname{Re}\Delta})$.
For $-1/2<\operatorname{Re}\Delta<0$, this growth is sublinear,
so Cauchy's estimates imply that $W$ is constant. We first
work in this range and then analytically continue the resulting
solution in $\Delta$. The constant is nonzero for a nontrivial
profile: on the real line $W=a^2w'$, while $a$ is nonvanishing
and $w$ is not constant.

On the real line, retaining the subtraction used to define
$\phi$, we have (up to $z$ independent factor)
\begin{equation}
    W(z) = a^2 w'(z) =2\pi i \Delta \exp\left(\frac{iz^2}{2} -\int_\R \frac{2\Delta S_\Delta(s)}{s-z-i0}ds + 2\pi i \Delta S_\Delta(z)\right) S'_\Delta(z).
\end{equation}
Taking its logarithmic derivative gives
\begin{equation}\label{rho1}
    \varrho'(z)=-iz\varrho(z)
       +\varrho(z)\left(R(z+i0)+R(z-i0)\right),
\end{equation}
where we have introduced the local density and its resolvent,
\begin{equation}
    \varrho(z)=\Delta S'_\Delta(z),\qquad
    R(z)=\int_{\R}\frac{\varrho(s)\,ds}{s-z}.
\end{equation}
Notice that \(\varrho(z)\) should not be confused with the original occupation profile \(\rho(k)\). 
Multiplying \eqref{rho1} by $1/(s-z)$, integrating over its
real argument $s$, and using $\int_{\R}\varrho(s)ds=\Delta$, we
obtain
\begin{equation}\label{Ricc}
    R'(z)+izR(z)+i\Delta-R(z)^2=0.
\end{equation}
This is a Riccati equation of the type that arises for Gaussian
ensemble resolvents (see, for example, \cite{EynardKimuraRibault2015}).
Moreover, as we show below, after the rotation $z=e^{-i\pi/4}s$
and a rescaling of the resolvent by $\Delta$, Eq.~\eqref{Ricc}
becomes Eq.~(12) of Ref.~\cite{AllezBouchaudGuionnet2012} upon
identifying $c=-\Delta$. In principle, this correspondence is
sufficient to determine $C_0$, \[
C_0=\frac{\Delta}{2}\ln(2\pi)+\ln G(1-\Delta)
\] as we demonstrate in the next section.
Nevertheless, it is useful to present an explicit solution of
Eq.~\eqref{Ricc}. To this end, we set
\begin{equation}
    R(z)=\frac{iz}{2}-\frac{u'(z)}{u(z)}.
\end{equation}
This leads to Weber's equation
\begin{equation}
    u''(z)+\left(\frac{1+2\Delta}{2i}
    +\frac{z^2}{4}\right)u(z)=0.
\end{equation}
Its general solution can be expressed in terms of
parabolic-cylinder functions:
\begin{equation}
    u(z)=c_1D_{-\Delta-1}\!\left(e^{i\pi/4}z\right)
         +c_2D_{\Delta}\!\left(e^{3i\pi/4}z\right).
\end{equation}
The matching condition
\begin{equation}
    R(z)=-\frac{\Delta}{z}+O(z^{-3}),
    \qquad z\to\infty,
\end{equation}
understood in closed subsectors of the respective half-planes,
selects, up to nonzero multiplicative constants,
\begin{equation}
    u_+(z)=D_\Delta\!\left(e^{-i\pi/4}z\right),
    \qquad
    u_-(z)=D_\Delta\!\left(-e^{-i\pi/4}z\right),
\end{equation}
and sets
\begin{equation}
    R(z)=
    \begin{cases}
        \displaystyle \frac{iz}{2}-\frac{u_+'(z)}{u_+(z)},
            & \operatorname{Im}z>0,\\[6pt]
        \displaystyle \frac{iz}{2}-\frac{u_-'(z)}{u_-(z)},
            & \operatorname{Im}z<0.
    \end{cases}
\end{equation}
Indeed, these solutions have the asymptotic behaviour
$u_\pm(z)\sim e^{iz^2/4}(\pm e^{-i\pi/4}z)^\Delta$
in the corresponding sectors.

The connection formula
\begin{equation}
    D_\Delta(w)
    =e^{-i\pi\Delta}D_\Delta(-w)
     +\frac{\sqrt{2\pi}\,e^{-i\pi(\Delta+1)/2}}{\Gamma(-\Delta)}
       D_{-\Delta-1}(iw)
\end{equation}
gives, with this normalization,
\begin{equation}
    c_1=
    \begin{cases}
        \displaystyle
        \frac{\sqrt{2\pi}\,e^{-i\pi(\Delta+1)/2}}{\Gamma(-\Delta)},
            & \operatorname{Im}z>0,\\[6pt]
        0,  & \operatorname{Im}z<0,
    \end{cases}
    \qquad
    c_2=
    \begin{cases}
        e^{-i\pi\Delta}, & \operatorname{Im}z>0,\\
        1,              & \operatorname{Im}z<0.
    \end{cases}
\end{equation}

Finally, the density follows from the Sokhotski--Plemelj jump
formula. Since
\begin{equation}
    u_+(z)u_-'(z)-u_+'(z)u_-(z)
    =\frac{\sqrt{2\pi}\,e^{-i\pi/4}}{\Gamma(-\Delta)},
\end{equation}
we find, for $z\in\mathbb{R}$,
\begin{multline}
    \varrho(z)
    =\frac{R(z+i0)-R(z-i0)}{2\pi i}
      =\frac{1}{2\pi i}
       \left(\frac{u_-'(z)}{u_-(z)}-\frac{u_+'(z)}{u_+(z)}
             \right)\\
    =-\frac{e^{i\pi/4}}
      {\sqrt{2\pi}\,\Gamma(-\Delta)\,
       D_\Delta(e^{-i\pi/4}z)D_\Delta(-e^{-i\pi/4}z)}.
\end{multline}
The regulator $S_\Delta(z)$ follows by integrating this expression
for $\varrho(z)$ and imposing boundary condition $S_\Delta(-\infty)=0$. An alternative
derivation that gives $S_\Delta$ directly is presented in
Appendix~\ref{appB}. 
Up to a contour rotation and rescaling, our result for $\varrho$
coincides with that of Ref.~\cite{AllezBouchaudGuionnet2012}.
More precisely, set $z=e^{-i\pi/4}s$ and define
\begin{equation}\label{prho}
    p(s)ds=\frac{1}{\Delta}\varrho(z)dz.
\end{equation}
For real $\Delta<0$, $p(s)$ agrees with Eq.~(18) of
Ref.~\cite{AllezBouchaudGuionnet2012} upon setting $c=-\Delta$.
We examine this random-matrix connection in more detail in the
next section, where we use the solution for $S_\Delta$ to evaluate
the constant $C_0$ in \eqref{c0}.

\section{The constant \(C_0\) and its random-matrix interpretation}
\label{sec:rmt}

In this section we use the relation between $\varrho(z)$ and the
high-temperature Gaussian $\beta$ ensemble. For real $\Delta<0$,
the density $p(s)$ obtained by the rotation $z=e^{-i\pi/4}s$ in
\eqref{prho} is the limiting ensemble density with $c=-\Delta$.
The partition function can be computed exactly, while its
thermodynamic limit is governed by a variational problem involving
the integrals needed to evaluate $C_0$.

For the analytic continuation, we write the real-line logarithm as
$\ln|s|=\frac12\ln s^2$ and follow its branch continuously along
the contour rotation. This is the logarithmic primitive of the
Cauchy kernel appearing in Appendix~\ref{appA}, and the same
continuation is inherited by \eqref{c0}. The clockwise rotation
shifts the logarithmic kernel by $-i\pi/4$. Since
$\int_{\mathbb R}p(s)\,ds=1$, this generates a contribution
$-i\pi\Delta^2/4$, which cancels the explicit
$i\pi\Delta^2/4$ in \eqref{c0}. Thus, in terms of the normalized
density $p(s)$, we obtain
\begin{equation}\label{C00}
    C_0 = \Delta\int_{\mathbb R}\frac{s^2}{2}p(s)\,ds
    +\Delta^2\iint_{\mathbb R^2}p(s)p(s')\ln|s-s'|\,ds\,ds'.
\end{equation}

The partition function of the Gaussian $\beta$ ensemble is
\begin{equation}
 Z_{N,\beta}
 =
 \int_{\mathbb R^N}
 \exp\!\left(-\frac12\sum_{j=1}^N s_j^2\right)
 \prod_{i<j}|s_i-s_j|^\beta
 \prod_{j=1}^N ds_j .
 \label{eq:gaussian-beta-Z}
\end{equation}
It has the exact expression~\cite{Mehta2004}
\begin{equation}\label{ZN}
    Z_{N,\beta}
 =(2\pi)^{N/2}
 \prod_{j=1}^N
 \frac{\Gamma(1+j\beta/2)}{\Gamma(1+\beta/2)}.
\end{equation}
We take the high-temperature limit
\begin{equation}
 \beta=\frac{2c}{N},
 \qquad N\to\infty,
 \qquad c=\text{fixed},
 \label{eq:high-temperature-scaling}
\end{equation}
and define the free energy per particle by
\begin{equation}
     F_*(c)
 =-\lim_{N\to\infty}\frac1N\ln Z_{N,2c/N}.
\end{equation}
From \eqref{ZN},
\begin{equation}
 \frac1N\ln Z_{N,2c/N}
 =\frac12\ln(2\pi)
 +\frac1N\sum_{j=1}^N
 \ln\Gamma\!\left(1+\frac{jc}{N}\right)
 -\ln\Gamma\!\left(1+\frac cN\right).
\end{equation}
The sum becomes a Riemann integral as $N\to\infty$, giving
\begin{equation}\label{criticalvalue}
 F_*(c)
 =-\frac12\ln(2\pi)
 -\frac1c\int_0^c\ln\Gamma(1+a)\,da .
\end{equation}

The same free energy has a variational representation,
$F_*(c)=\min_p\mathscr F_c[p]$, over positive, normalized densities
$\int_{\mathbb R}p(s)\,ds=1$ \cite{NakanoTrinh2020}. The functional is
\begin{equation}\label{fcp}
    \mathscr F_c[p]
 =
 \int_{\mathbb R}p(s)\ln p(s)\,ds
 +\frac12\int_{\mathbb R}s^2p(s)\,ds
 -c\iint_{\mathbb R^2}p(s)p(v)\ln|s-v|\,ds\,dv .
\end{equation}
Varying this functional gives an equation for its minimizer $p_c$:
\begin{equation}
   \ln p_c(s)+\frac12s^2
   -2c\int_{\mathbb R}p_c(v)\ln|s-v|\,dv+\Lambda=0,
\end{equation}
where $\Lambda$ enforces normalization.\footnote{The multiplier can also be determined from the partition-function ratio (see discussion after Eq.~(9) in \cite{NakanoTrinh2020}). Namely, the limit $N\to \infty$, 
$\displaystyle
\frac{Z_{N,\beta_N}}{Z_{N-1,\beta_N}}
\to
\int_{\mathbb R}
\exp\!\left[-\frac{x^2}{2}
+2c\int_{\mathbb R}p_c(y)\ln|x-y|\,dy\right]dx
=e^\Lambda$.
Therefore using \eqref{ZN} with $\beta_N=2c/N$ we obtain
$e^\Lambda=\sqrt{2\pi}\Gamma(1+c)$.}
Differentiating with respect to $s$ gives the rotated and rescaled form of
\eqref{rho1}, with $c=-\Delta$.

We can extract the two integrals in \eqref{C00} without evaluating $p_c$
explicitly. First, $c$ enters \eqref{fcp} explicitly only through the
logarithmic interaction. Differentiating the functional at its minimizer
therefore gives
\begin{equation}
 \frac{\partial F_*(c)}{\partial c}
 =\left.\frac{\partial \mathscr F_c[p]}{\partial c}
 \right|_{p=p_c}
 =-\iint_{\mathbb R^2}p_c(s)p_c(v)\ln|s-v|\,ds\,dv .
 \label{eq:envelope}
\end{equation}
Combining this relation with \eqref{criticalvalue} and the standard
Barnes-$G$ integral identity \cite{DLMFBarnesG}, we obtain
\begin{equation}
 c^2\iint_{\mathbb R^2}p_c(s)p_c(v)\ln|s-v|\,ds\,dv
 =\ln G(1+c)
 -\frac c2\ln(2\pi)
 +\frac{c(c+1)}2 .
 \label{eq:Ic-Barnes}
\end{equation}

To find the quadratic moment, consider the rescaled density
$p_\lambda(s)=\lambda^{-1}p_c(s/\lambda)$, which remains normalized.
Under this rescaling,
\begin{equation}
  \mathscr F_c[p_\lambda]
  =\mathscr F_c[p_c]-(1+c)\ln\lambda
  +\frac{\lambda^2-1}{2}\int_{\mathbb R}s^2p_c(s)\,ds.
\end{equation}
Since $p_c$ minimizes the functional, the derivative at $\lambda=1$
vanishes. Thus
\begin{equation}
    \int_{\mathbb R}s^2p_c(s)\,ds=1+c.
\end{equation}
Setting $c=-\Delta$ in \eqref{C00}, the quadratic term cancels the
polynomial term in \eqref{eq:Ic-Barnes}. 
On the branch obtained by analytic continuation from
$-1/2<\Delta<0$, these identities yield the prefactor
\eqref{prefactor}.

\section{Relation to the standard Riemann--Hilbert problem}
\label{sec:rhp}

The appearance of parabolic-cylinder functions in the saddle point
regularization parallels their role in the Riemann--Hilbert analysis
\cite{Minin2026}. In fact, under suitable nonvanishing and regularity
assumptions, solving the full Riemann--Hilbert problem is equivalent
to determining an effective phase for which $K=\mathcal{V}$ in the
notation of Sec.~\ref{sec:eff}. The original kernel can then be
written in the effective form, and its determinant is given by
\eqref{eq:eff-functional} for the reconstructed phase.

To see this, consider the standard Riemann--Hilbert problem (RHP)
associated with $K$ \cite{Korepin1993}; see also
Appendix~\ref{appA}. We seek a $2\times 2$ matrix $\chi(q)$,
holomorphic off the real line, normalized by
$\chi(q)=\mathds{1}+O(1/q)$ as $q\to\infty$, and satisfying
\begin{equation}\label{rhp02}
    \chi_-(q) = \chi_+(q)\left(\begin{array}{cc}
    	1+2\pi ie_-(q)e_+(q)  & 2\pi i e_-(q)^2\\
    	-2\pi i e_+(q)^2& 1-2\pi i e_-(q)e_+(q)
    \end{array}\right) ,\qquad q \in \mathbb{R}.
\end{equation}
Here $\chi_\pm(q)\equiv\chi(q\pm i0)$ denote the boundary values
from the upper and lower half-planes, respectively.\footnote{The
subscripts on $\chi$ label boundary values and are independent of
those on the functions $e_\pm$ defined in \eqref{eq:epem}.}
Usually, $\chi$ is used to reconstruct the resolvent of $K$.
Here we instead use its matrix elements to express the ratio
$e_+/e_-$. The second row of the jump relation gives
\begin{align}\label{5.2}
     (\chi_-)_{21} &=  \left(1+2\pi i e_-e_+ \right)(\chi_+)_{21}
     - 2\pi i (\chi_+)_{22}e_+^2, \\ \label{5.3}
     (\chi_-)_{22} &=  2\pi i (\chi_+)_{21} e_-^2
     +\left(1-2\pi ie_-e_+ \right)(\chi_+)_{22}.
\end{align}
Assume that $\chi_{22}$ has no zeros in either half-plane or in
its boundary values on the real line. We may then introduce
\begin{equation}
    \psi(q) \equiv \frac{(\chi_-(q))_{22} }{(\chi_+(q))_{22}},
    \quad q\in \R; \qquad
    \phi(q) = \frac{\chi_{21}(q)}{\chi_{22}(q)}.
\end{equation}
Equations \eqref{5.3} and \eqref{5.2} imply
\begin{equation}
    \frac{e_+(q)}{e_-(q)}
    = \phi_+(q) + \frac{1-\psi(q)}{2\pi i e_-^2(q)},
    \qquad
    \phi_+ - \phi_-
    = \frac{\psi(q)+1/\psi(q) -2}{2\pi i e_-^2(q)}.
\end{equation}
Since $\phi(q)=O(1/q)$ at infinity, the second relation determines
its Cauchy representation, provided the jump has sufficient decay:
\begin{equation}
    \phi(q) = \int \frac{dk}{(2\pi i)^2 }
    \frac{\psi(k)+1/\psi(k) -2}{e_-^2(k) (k-q)}.
\end{equation}
Substituting its upper boundary value into the first relation gives
\begin{equation}
    \frac{e_+(q)}{e_-(q)}
    =\frac{1}{4\pi^2} \int \frac{dk}{e_-(k)^2} \left(
    \frac{1-\psi(k)}{k-q+i0}
    + \frac{1-1/\psi(k)}{k-q-i0}
    \right).
\end{equation}
With $v_-=e_-$, this expression coincides with
\eqref{eq:epem1} after the identification
\begin{equation}
 \psi(k) = e^{2\pi i \eta(k)}.
\end{equation}
The logarithmic branch defining $\eta$ is fixed by choosing the
analytic logarithms of $\chi_{22}$ in the two half-planes to
vanish at infinity.

Conversely, Appendix~\ref{appA} constructs the normalized RHP
solution from an admissible effective phase and weight. Thus,
within this class, exact kernel matching and the RHP give
equivalent descriptions of the determinant. The phase reconstructed
in this way is generally $x,t$-dependent; the outer branches and
local profile obtained in Secs.~\ref{sec:eff} and
\ref{sec:sewing} describe its asymptotic form. This identification
also provides a direct route to comparing the sewing conditions
\eqref{cond1} with the saddle-point model RHP constructed in
Appendix~E of Ref.~\cite{Minin2026}.

\section{Summary and outlook}
\label{summ}

We have extended the effective form factor construction to a
time-dependent generalized sine kernel with a momentum-dependent
phase. Matching the original and effective kernels determines
the auxiliary weight and two outer branches of the effective
phase. The main advance is an explicit local profile joining
these branches across their discontinuity within a region of
width $t^{-1/2}$. After a contour rotation, the derivative of
this profile becomes the limiting density of the high-temperature
Gaussian $\beta$ ensemble \cite{AllezBouchaudGuionnet2012}.
Its exact partition function determines the jump-dependent
constant and fixes the finite prefactor of the effective
determinant asymptotic, in agreement with the Riemann--Hilbert
result of Ref.~\cite{Minin2026}.

We have also established a direct connection between the EFF
construction and the standard matrix Riemann--Hilbert problem.
Under the stated nonvanishing and regularity assumptions, the
effective phase is determined by the ratio of the boundary
values of $\chi_{22}$, and exact kernel matching becomes a
reformulation of the full RHP. The local EFF construction can
therefore be viewed as an asymptotic description of this exact
phase. This connection clarifies the common origin of the
parabolic-cylinder functions appearing in the two approaches.

The hard-rod application shows that the outer asymptotics
already recover the Euler-scale density correlation predicted
by generalized hydrodynamics, without requiring the exact
finite prefactor \cite{GamayunPanfil2026}.

Since the jump-dependent contribution is local and depends
only on the discontinuity, we expect the same local factor
in other systems with a nondegenerate quadratic saddle and
the same local matching structure. In particular, this should
apply to lattice systems \cite{ZhuravlevEtAl2022}, with a
separate local analysis at each of the two stationary points
of the dispersion. Another possible application concerns the
asymptotics of tau functions of classical integrable hierarchies.

We also expect the method to extend to Fredholm determinants
with corrections of finite rank. These corrections can be
treated using the explicitly known effective resolvent, or
through direct effective form factor representations of
integrals involving the Fredholm determinant
\cite{GamayunIorgovZhuravlev2021}. Such extensions may also be
useful for interacting systems, where dual-field representations
provide a possible framework for formulating the corresponding
effective matching problem.

A different extension concerns discontinuities already present
in the state, such as Fermi edges at zero temperature. These
require their own local matching analysis. Our preliminary
analysis suggests an analogous structure involving Tricomi
confluent hypergeometric functions and partition functions of
high-temperature Laguerre ensembles, in place of
parabolic-cylinder functions and the Gaussian ensemble.
We leave a detailed treatment of this problem to a separate
paper.

A fully rigorous derivation within the effective form factor
approach still requires several steps. The analytic and growth
assumptions used to select the local Weber solution must be
established, and the contour rotation and subsequent analytic
continuation in $\Delta$ must be justified. This includes
controlling the oscillatory integrals and verifying that no
poles are crossed in the initial domain of the deformation.

For a proof based on the factorization
\eqref{eq:kernel-factorization}, one must establish
\begin{equation}
    \det\left(1+(1-\hat{\mathcal R})\hat{\mathcal W}\right)
    =1+o(1)
\end{equation}
as $t\to\infty$ at fixed $x/t$, with $\hat{\mathcal W}$ denoting
the remainder associated with the asymptotically matched
effective kernel. The required estimate must account for the
action of the effective resolvent on this remainder.
Alternatively, the exact RHP reformulation suggests controlling
the difference between the determinant functional evaluated
on the exact phase and on its matched asymptotic approximation.
Together with the local analysis, such global estimates would
provide an independent proof of the full Fredholm determinant
asymptotic.

\section*{Acknowledgements}

I am grateful to Mi{\l}osz Panfil and Yuri Zhuravlev for their
careful reading of the manuscript and numerous useful suggestions.
I would also like to thank the organizers of the RAQIS 2026
conference, where many fruitful discussions inspired this paper.
I acknowledge support from Cognia AI.

\appendix

\section{Fredholm determinant for the effective form factors}
\label{appA}

In this appendix we derive the explicit expression
\eqref{eq:eff-functional} for the Fredholm determinant of the kernel
\eqref{tt}. We follow the construction of
Ref.~\cite{gamayun_finite-temperature_2025}, adapted to the real line
as in Ref.~\cite{ChernowitzGamayun2022}. The key step is to compute
explicitly the resolvent $\mathcal{R}$ defined by
\begin{equation}
    (1+\mathcal{V})(1-\mathcal{R}) =1
\end{equation}
We introduce the standard vector notation \cite{Korepin1993}
\begin{equation}
	|V(q)\rangle = \left(\begin{array}{c}
		v_-(q) \\ -v_+(q)
	\end{array}\right),\qquad \langle V(q)|  = (v_+(q),v_-(q)),
\end{equation}
\begin{equation}
	|F(q)\rangle = \left(\begin{array}{c}
		f_-(q) \\ -f_+(q)
	\end{array}\right),\qquad \langle F(q)|  = (f_+(q),f_-(q)).
\end{equation}
The kernels then take the form
$\mathcal{V}(q,p) = \frac{\langle V(q)| V(p)\rangle}{q-p}$ and
$\mathcal{R}(q,p) = \frac{\langle F(q)| F(p)\rangle}{q-p}$.

Consider the matrix-valued function
\begin{equation}\label{chiQ}
    \chi(q) = \mathds{1}_{2\times 2} - \int_\R  \frac{|F(k)\rangle \langle V(k)|}{k-q}dk
\end{equation}
It is holomorphic off the real line. Denoting its boundary values
from the upper and lower half-planes by
$\chi_\pm(q)\equiv\chi(q\pm i0)$, we obtain the jump relation
\begin{equation}\label{rhp2}
    \chi_-(q) = \chi_+(q)J_\chi ,\qquad q \in \mathbb{R}
\end{equation}
with
\begin{equation}
    J_\chi  =  \mathds{1} + 2\pi i |V(q)\rangle\langle V(q)|=
    	\left(\begin{array}{cc}
    	1  &0\\
    	0& 1
    \end{array}\right)+  2\pi i\left(\begin{array}{cc}
    	v_-v_+  & v_-^2\\
    	-v_+^2& -v_-v_+
    \end{array}\right).
\end{equation}
Once $\chi$ is found from the Riemann--Hilbert problem (RHP)
\eqref{rhp2}, the resolvent is recovered through
$|F(q)\rangle = \chi_\pm(q)|V(q)\rangle$ for $q\in\R$.
The key observation is that this RHP can be solved exactly by a
sequence of triangular and diagonal transformations.

The first transformation is
\begin{equation}
	\chi = \Phi \left(\begin{array}{cc}
		1 & 0 \\
		\varphi& 1
	\end{array}\right)
\end{equation}
where
\begin{equation}
    \varphi(q) =\int_\R \frac{dk}{2\pi i} \frac{e^{-i\mathfrak{p}(k)x}\sigma(k)}{k-q}.
\end{equation}
To remove the Cauchy integrals from the jump matrix, we choose
\begin{equation}
    \sigma(q) = \frac{e^{2\pi i \eta(q)}+e^{-2\pi i \eta(q)}-2}{2\pi i r(q)}
\end{equation}
The resulting jump is upper triangular
\begin{equation}
    \Phi_- = \Phi_+ J_\Phi,\qquad J_\Phi = \left(
    \begin{array}{cc}
    	e^{-2\pi i \eta} & 2\pi i v_-^2\\
    	0 & e^{2\pi i \eta}
    \end{array}
    \right).
\end{equation}
The next conjugation sets the diagonal entries of the jump to unity
\begin{equation}
    \Phi = \Psi \left(\begin{array}{cc}
    	e^{-\Omega} & 0\\
    	0 & e^{\Omega}
    \end{array}\right),\qquad
\Omega(q) = - \int_\R \frac{\eta(k)dk}{k-q}.
\end{equation}
The transformed jump relation is $\Psi_-=\Psi_+J_\Psi$, with
\begin{equation}
 J_\Psi = \left(
    \begin{array}{cc}
    	1 & 2\pi i v_-^2e^{-\omega}\\
    	0 & 1
    \end{array}
    \right),\qquad \omega(q)\equiv \Omega_+(q)+\Omega_-(q) = - 2{\rm v.p.} \int_\R  \frac{\eta(k)dk}{k-q}.
\end{equation}
This problem has the explicit solution
\begin{equation}
    \Psi(q) =   \left(\begin{array}{cc}
    	1&  b(q)\\
    	0 & 1
    \end{array}\right)  ,\qquad b(q)= -\int_\R \frac{v_-^2(k)e^{-\Omega_+(k)-\Omega_-(k)}}{k-q} dk
\end{equation}
Combining the transformations gives
\begin{equation}\label{chi2}
\chi(q) =
 \left(\begin{array}{cc}
    	1&  b(q)\\
    	0 & 1
    \end{array}\right)   \left(\begin{array}{cc}
    	e^{-\Omega(q)} & 0\\
    	0 & e^{\Omega(q)}
    \end{array}\right)
    \left(\begin{array}{cc}
		1 & 0 \\
		\varphi(q)& 1
	\end{array}\right).
\end{equation}
This solution has the required normalization as $q\to\infty$.

To evaluate the determinant, we vary $\eta(k)$ while keeping $r(k)$
fixed. The corresponding kernel variation is
\begin{equation}
    \delta_\eta V(q,p) = \frac{v_-(q)v_-(p)}{2\pi i} \int_\R \frac{dk}{v_-^2(k)} \left[
    \frac{e^{2\pi i \eta(k)}\delta \eta(k)}{(k-q+i0)(k-p+i0)}-
    \frac{e^{-2\pi i \eta(k)}\delta \eta(k)}{(k-q-i0)(k-p-i0)}
    \right]
\end{equation}
and the logarithmic variation of the determinant reads
\begin{equation}
    \delta_\eta \ln \det(1+\mathcal{V}) = {\rm Tr} (1-\mathcal{R})\delta_\eta V
\end{equation}
The Cauchy representation \eqref{chiQ} gives the identity
(cf.~Appendix A of Ref.~\cite{gamayun_finite-temperature_2025})
\begin{equation}
    \iint_{\mathbb R^2} \frac{v_-(q) (1-\mathcal{R})(q,p) v_-(p)}{(k_1-q)(k_2-p)}  dq \,dp = \frac{\chi_{22}(k_1)\chi_{12}(k_2)-\chi_{12}(k_1)\chi_{22}(k_2)}{k_1-k_2},
\end{equation}
Substituting the explicit solution \eqref{chi2}, we obtain
\begin{equation}
    \iint_{\mathbb R^2} \frac{v_-(q) (1-\mathcal{R})(q,p) v_-(p)}{(k_1-q)(k_2-p)}  dq \,dp = -e^{\Omega(k_1)+\Omega(k_2)}\frac{b(k_1)-b(k_2)}{k_1-k_2}.
\end{equation}
Taking the limit $k_1\to k_2$ then yields
\begin{equation}\label{eq:Afix-trace}
    {\rm Tr} (1-\mathcal{R}) \delta_\eta V = \int_\R \frac{dk}{v_-^2(k)} \frac{\delta \eta(k)}{2\pi i}
    \left(
    e^{2\Omega_-(k) - 2\pi i \eta(k)} b'_-(k) - e^{2\Omega_+(k) + 2\pi i \eta(k)} b'_+(k)
    \right)
\end{equation}
The boundary values of $b(k)$ and $\Omega(k)$ satisfy
\begin{align}\label{eq:Afix-bjump}
    b_-(k)- b_+(k) =&  \, 2\pi i v_-^2(k) e^{-\omega(k)}\\
    \Omega_-(k) - \Omega_+(k) = & \,2\pi i \eta(k).
\end{align}
Since $\omega=\Omega_++\Omega_-$, we have
$e^{2\Omega_--2\pi i\eta}
=e^{2\Omega_++2\pi i\eta}=e^\omega$.
Differentiating the jump relation \eqref{eq:Afix-bjump} therefore
simplifies \eqref{eq:Afix-trace} to
\begin{multline}
    \delta_\eta\ln\tau
= \int_{\mathbb R}\frac{\delta\eta(k)e^{\omega(k)}}{2\pi i\,v_-^2(k)}
\bigl[b_-'(k)-b_+'(k)\bigr] = \int_{\mathbb R}\delta\eta(k)
\left[\frac{(v_-^2)'(k)}{v_-^2(k)}-\omega'(k)\right]dk  \\
=\int_{\mathbb R}\delta\eta(k)
\left[i(x-tk)+\frac{r'(k)}{r(k)}-\omega'(k)\right]dk\\ =\int_{\mathbb R}\delta\eta(k)
\left[i(x-tk)+\frac{r'(k)}{r(k)}+2{\rm v.p. }\int_\R \frac{\eta'(q)dq}{q-k}\right]dk.
\end{multline}
This coincides with the variation of the functional on the
right-hand side of \eqref{eq:eff-functional}. Denoting its exponential
by $\tau[\eta]$, we have $\tau[\eta=0]=1$ and
$\det(1+\mathcal{V})\Big|_{\eta=0}=1$. This normalization fixes the
integration constant and completes the proof of
\eqref{eq:eff-functional}.

\section{Alternative solution}
\label{appB}

The sewing conditions \eqref{cond1} can equivalently be expressed
in terms of Fourier modes. For $\gamma>0$, they read
\begin{equation}\label{fff}
    \int\limits_{-\infty}^\infty(w(s)-1)e^{is^2/2+i\gamma s} ds =0,\qquad     \int\limits_{-\infty}^\infty(e^{2\pi i \Delta }w(s)^{-1}-1)e^{is^2/2-i\gamma s}ds =0.
\end{equation}

The substitution $s\to-s$ suggests seeking a solution satisfying
$w(s)w(-s)=e^{2\pi i\Delta}$. We therefore use the ansatz
$w(s)=e^{\pi i\Delta}U(-s)/U(s)$.
We seek a function $U(s)$ that extends analytically into the upper
half-plane and has the positive Fourier-mode representation
\begin{equation}\label{U}
    U(s) = \int\limits_0^\infty  e^{i\lambda s} u(\lambda) d\lambda.
\end{equation}
Introduce
$f(s)=e^{is^2/2}(e^{i\pi\Delta}U(-s)-U(s))$.
This is the product of $U(s)$ and $(w(s)-1)e^{is^2/2}$, both of
which contain only positive Fourier modes. Its Fourier transform
therefore vanishes for $\gamma>0$. Evaluating the transform using
\eqref{U} gives
\begin{equation}
 \int\limits_{-\infty}^\infty f(s) e^{i\gamma s} ds =0 \quad\Leftrightarrow\quad \int\limits_0^\infty (e^{i\pi \Delta + i\lambda \gamma}-e^{-i\lambda\gamma})e^{-i\lambda^2/2}u(\lambda)d\lambda=0
\end{equation}
Using the integral representation of the gamma function
\footnote{\url{http://dlmf.nist.gov/5.9.E1}}
\begin{equation}
\int_{0}^{\infty} p^{\alpha-1} e^{\pm i r p}\,dp
=
\Gamma(\alpha)\,r^{-\alpha}
e^{\pm i\pi\alpha/2},
\qquad
r>0,\quad 0<\operatorname{Re}\alpha<1,
\end{equation}
we find that $u(\lambda)=e^{i\lambda^2/2}\lambda^{-\Delta-1}$
satisfies this condition, initially for
$-1<\operatorname{Re}\Delta<0$.
Analytic continuation can be implemented by replacing the contour
in \eqref{U} with a Hankel contour. The resulting integral expresses
$U$ in terms of parabolic-cylinder functions. The common even
Gaussian factor and constant normalization cancel in the ratio
$U(-s)/U(s)$, giving
\begin{equation}
    w_\Delta(z)=e^{i \pi \Delta}
\frac{D_\Delta(-e^{-i\pi/4}  z)}{D_\Delta(e^{-i\pi/4} z)},\qquad
S_\Delta(z)=
\frac{1}{2}+\frac{1}{2\pi i \Delta}\ln\frac{D_\Delta(-e^{-i\pi/4} z)}{D_\Delta(e^{-i\pi/4} z)}.
\end{equation}
These expressions extend by analytic continuation in $\Delta$,
with the logarithmic branch fixed by continuation from zero jump.

\nolinenumbers
\begingroup
\raggedright
\bibliography{effective_form_factors_weber}
\endgroup
\end{document}